# Boron Nitride Nanosheets Improve Sensitivity and Reusability of Surface Enhanced Raman Spectroscopy

Qiran Cai,[1] Srikanth Mateti,[1] Wenrong Yang,[2] Rob Jones,[3] Kenji Watanabe,[4] Takashi Taniguchi,[4] Shaoming Huang,[5] Ying Chen[1]* and Lu Hua Li[1]*

1. Institute for Frontier Materials, Deakin University, Geelong Waurn Ponds Campus, VIC 3216, Australia

2. Centre for Chemistry and Biotechnology, School of Life and Environmental Sciences, Deakin University, Geelong Waurn Ponds Campus, VIC 3216, Australia

3. Department of Physics, La Trobe University, Bundoora, 3086, VIC (Australia)

4. National Institute for Materials Science, Namiki 1-1, Tsukuba, Ibaraki 305-0044, Japan

5. Nanomaterials and Chemistry Key Laboratory, Wenzhou University, Wenzhou 325027, China.

**Abstract: Surface enhanced Raman spectroscopy (SERS) is a useful multidisciplinary analytic technique. However, it is still a challenge to produce SERS substrates that are highly sensitive, reproducible, stable, reusable, and scalable. Here, we demonstrate that atomically thin boron nitride (BN) nanosheets have many unique and desirable properties to help solve this challenge. The synergic effect of the atomic thickness, high flexibility, stronger surface**

**adsorption capability, electrical insulation, impermeability, high thermal and chemical stability of BN nanosheets can increase the Raman sensitivity by up to two orders, and in the meantime attain long-term stability and extraordinary reusability not achievable by other materials. These advances will greatly facilitate the wider use of SERS in many fields.**

Surface enhanced Raman spectroscopy (SERS) is one of a few analytical techniques of single-molecule sensitivity,[1] and has a wide range of applications in research and industry. In spite of decades of study, the production of highly-sensitive, homogeneous, reproducible, reusable, and cost-effective SERS substrates is still a big challenge. Two-dimensional (2D) nanomaterials which have many unique characteristics and properties provide new possibilities for SERS. For example, high Raman scattering has been observed from graphene.[2] When graphene is used to separate plasmonic metal particles and analyte molecules, both the signal and reproducibility of SERS could be improved.[3] Furthermore, graphene is highly impermeable, and could prevent the underlying silver (Ag) particles from oxidation at room temperature.[4] However, such protection is effective only in the short term,[5] and in the long run it actually speeds up oxidation due to galvanic corrosion.[6] In addition, graphene cannot solve the reusability challenge because it oxidizes at only 250 °C in air.[7]

Boron nitride (BN) nanosheets, atomically thin layers of hexagonal boron nitride (hBN), are another important member of 2D nanomaterials. BN nanosheets have mechanical strength, thermal conductivity, and impermeability similar to graphene, but are electrical insulators thermally stable up to 800 °C in air.[8] BN nanosheets could greatly contribute to SERS field. Take reusability as an example. It was found that thorough removal of adsorbed analyte molecules could, in most cases, be achieved only by heating treatments in oxygen or air but not by solvent washing.[9] Thus, to be reusable, SERS substrates need to withstand oxidation at high temperatures, which is especially difficult for oxidation-sensitive Ag nanoparticles. Traditional coatings, such as alumina ($Al_2O_3$), silica ($SiO_2$), and zinc oxide ($ZnO_2$) can increase the reusability of Ag nanoparticles,[10] but at the cost of dramatically reduced Raman enhancement: a 1.5 nm thick $Al_2O_3$ coating could reduce Raman enhancement by 75%.[11] Another problem of ceramic coatings is their low affinity for most molecules, which means less number of molecules available for analysis, and hence weaker SERS signals. Atomically thin BN nanosheets can solve these problems. However, the use of atomically thin BN in SERS has been much less explored.[9b, 12] More importantly, there still lack proof-of-concept studies to demonstrate the full potential of atomically thin BN in SERS application.

Here, we show that atomically thin BN nanosheets can improve the sensitivity, reproducibility, and reusability of SERS substrates. The high flexibility makes atomically thin BN wrinkled to closely follow the contours of the underlying nanoparticles, with minimum decay of localized surface plasmon induced electromagnetic fields. Atomically thin BN has a stronger surface adsorption capability, and hence is more efficient in adsorption of analyte molecules at extremely low concentrations, resulting in dramatically improved sensitivity. Atomically thin BN is electrically insulating, so it can eliminate undesirable charge transfer as well as photocatalytic decomposition of analyte molecules to enhance reproducibility and stability of SERS. More importantly, atomically thin BN can protect Ag nanoparticles from oxidation in air even at high temperatures due to its high thermal stability and excellent impermeability. That is, the BN nanosheet covered SERS substrates have an outstanding reusability without loss of Raman enhancement.

The high-quality BN nanosheets used in this study were mechanically exfoliated from hBN single crystals (see Supporting Information, Figure S2),[13] and rhodamine 6G (R6G), the most commonly used analyte for SERS analysis, was chosen as a model molecule. We first reveal the stronger surface adsorption capability of atomically thin BN. After the BN nanosheets of different thicknesses immersed in R6G aqueous solution of $10^{-6}$ M for 1 min, the atomic force microscopy (AFM) images (Figure 1a and b) show that the thickness increase due to R6G adsorption for all BN nanosheets was the same, i.e. ~0.9 nm, corresponding to a monolayer of lying-down R6G molecules, whose xanthene ring structures are parallel to BN surface due to strong $\pi$-$\pi$ interactions.[14] However, the amount of the adsorbed molecules decreased with the increase of BN's thickness. Such intriguing phenomenon can be better seen in Figure 1c, where the surface of all the BN nanosheets were adjusted to the same color (i.e. AFM height) so that their coverage of R6G could be directly compared. The bilayer (2L) BN nanosheet was almost completely covered by R6G molecules, whereas the surface of a bulk hBN (~50 nm thick) adsorbed much less molecules. The adsorption coverage estimated from the AFM images[15] is 90% on 2L BN, ~70% on 3-6L BN, and less than 40% on the bulk BN, respectively (Figure 1d). Thus, the thinner the BN nanosheet, the higher the surface adsorption. According to our recent investigation, such phenomenon can be ascribed to conformational change in atomically thin nanosheets, which increases adsorption energy and efficiency (see Supporting Information for Review ONLY). Conformational change often happens to soft adsorbates, such as macromolecules and biomolecules, and has been rarely observed on adsorbent. However, atomically thin BN nanosheets are highly flexible, so that adsorbent, i.e. atomically thin BN, instead of adsorbate, R6G, experienced conformational change in the current case. Bulk hBN crystals are not able to perform conformational change, so they have worse adsorption performance than atomically thin BN nanosheets.

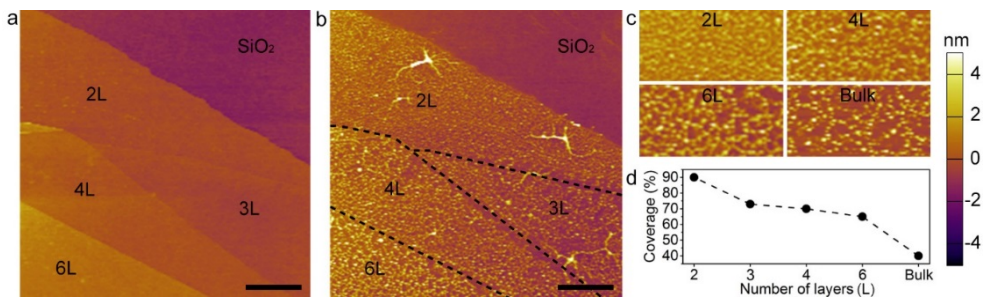

**Figure 1.** AFM images of BN nanosheets before (a) and after (b) R6G adsorption; (c) normalized AFM images with the surface of BN of different thicknesses adjusted to the same height to reveal the different coverage of adsorbed R6G molecules under the same adsorption condition; (d) estimated percentages of coverage v.s. BN thickness. Scale bars 2 μm.

The unique adsorption behavior and superior adsorption performance of atomically thin BN can improve the sensitivity of SERS. For this purpose, BN nanosheets were placed on top of plasmonic Ag nanoparticles, produced by thermal annealing of mechanically exfoliated BN nanosheets on sputtered Ag film on $SiO_2$/Si substrate (see Supporting Information for Experimental Details). As shown in Figure 2a and b, the flexible 2L BN wrinkled following the profile of the underlying nanoparticles; thicker (17L) BN is more rigid and deformed to a much smaller degree (Figure 2c and d). Their thicknesses (0.79 nm for 2L and 5.91 nm for 17L) were determined before annealing (Supporting Information, Figure S3). Figure 2b and d show representative AFM height traces and the corresponding diagrams. The atomically thin and wrinkled 2L BN should give rise to stronger Raman enhancement than the 17L BN because plasmon-induced electromagnetic fields (i.e. hot spots) diminish exponentially with distance. In other words, atomically thin BN can reserve the plasmonic hot spots. Note that the diameter and distribution of Ag nanoparticles with and without BN coverage are similar (see Supporting Information, Figure S4), and according to our previous study, the dielectric contant of BN nanosheets of differnt thicknesses is quite close.[16]

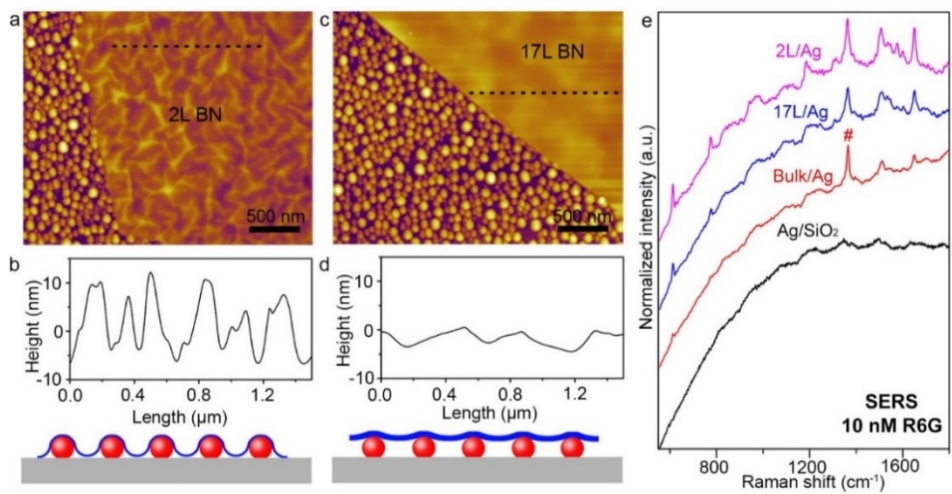

**Figure 2.** AFM images of Ag nanoparticles covered by 2L (a) and 17L (c) BN; (b) and (d) are the height traces and schematic diagrams corresponding to the dotted lines in (a) and (c); (e) SERS spectra of R6G (10 nM) from 2L, 17L and bulk BN, as well as the Ag/SiO₂ substrate by adsorption method, and # represents the Raman G band ($E_{2g}$ mode) of bulk hBN.

The sensitivity of the SERS substrates veiled by BN nanosheets of different thicknesses and a bare Ag substrate was tested by immersing in $10^{-8}$ M aqueous solution of R6G. As expected, the Raman features of R6G were most prominent from the 2L BN covered Ag nanoparticles (2L/Ag), and the signal attenuated with the increase of BN thickness (Figure 2e). The bare Ag nanoparticles on SiO₂/Si substrate (Ag/SiO₂), on the other hand, showed no Raman signature of R6G at either $10^{-8}$ or $10^{-7}$ M (Figure 2e and Supporting Information, Figure S5). That is, the 2L BN increased the sensitivity of the SERS substrates by more than two orders of magnitude. The much better enhancement from the atomically thin BN covered SERS substrate can be largely attributed to the aforementioned two reasons: (1) atomically thin BN nanosheets are more efficient in capture of analyte molecules, and (2) better reserved Raman hot spots due to their atomic thickness and wrinkles. Between these two causes, the first one played a decisive role because 1) the enhancement factors for these samples are close (see Supporting Information); 2) theoretically, the $Ag/SiO_2$ substrate should have the strongest electromagnetic field but showed no meaningful signal due to its low affinity for aromatic molecules, whereas Bulk/Ag still showed weak Raman peaks of R6G (Figure 2e). This was further confirmed by estimating the enhancement factors of the different SERS substrates (see Supporting Information). Distinct from graphene which shows strong Raman peaks, the Raman yield of atomically thin BN is so weak that its G band is not visible. BN nanosheets can also improve SERS reproducibility and stability because they act as dielectric separators to prevent undesired decomposition of analyte molecules caused by the photocatalysis of Ag nanoparticles.

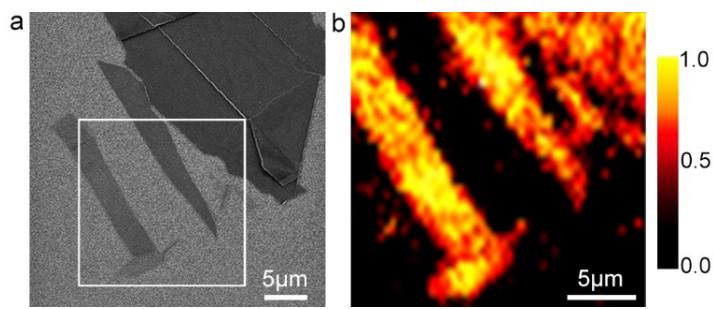

**Figure 3.** (a) Scanning electron microscopy (SEM) image of Ag particles covered by BN of different thicknesses; (b) the corresponding Raman mapping from the squared area in (a) after immersion in $10^{-6}$ M R6G solution for 1 h, where 0.0 means no intensity and 1.0 means the strongest measured Raman peak at 612 $cm^{-1}$.

The homogeneity and reproducibility of the SERS substrates were checked by Raman mapping, as shown in Figure 3. The contrast of the Raman intensity in the areas with and without BN coverage is striking: the $Ag/SiO_2$ areas are mostly in black, indicating much weaker Raman sensitivity; whereas the signals from BN/Ag areas are several times stronger. In addition, the Raman enhancement from the BN areas is relatively homogeneous. The weaker Raman signals around the edges of the BN nanosheets should be due to the average of the signals from BN/Ag and $Ag/SiO_2$.

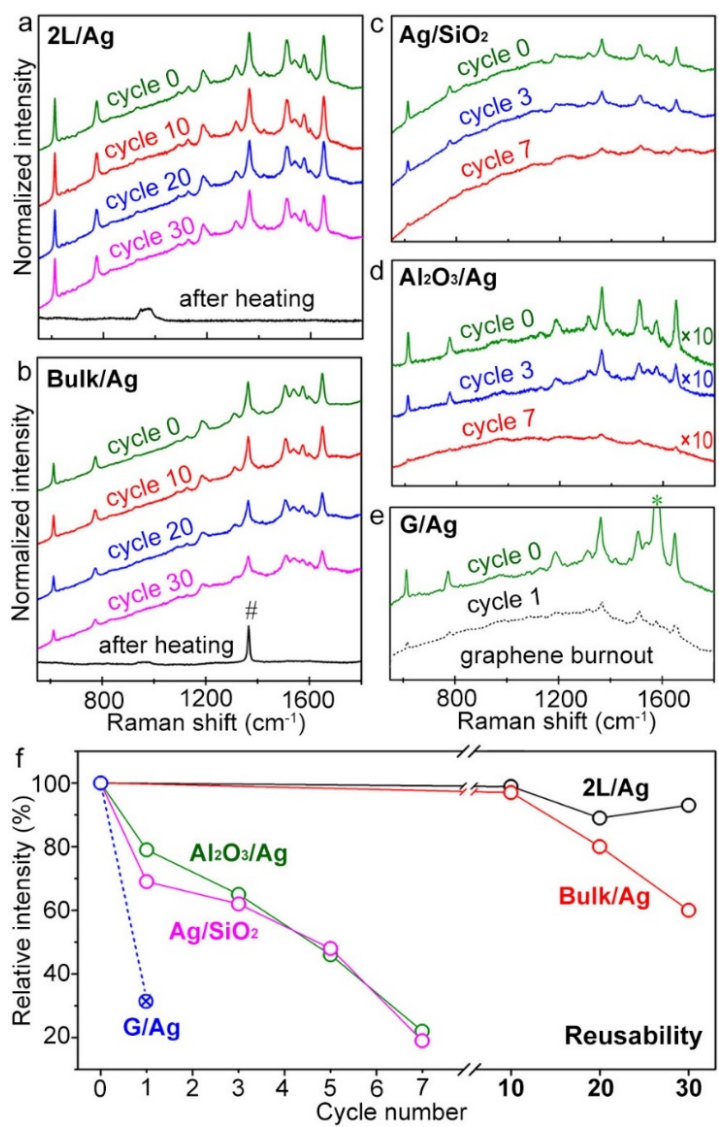

**Figure 4.** Reusability tests of different SERS substrates: (a) 2L BN covered Ag nanoparticles (2L/Ag) for 30 cycles; (b) bulk BN on Ag nanoparticles (Bulk/Ag) for 30 cycles, where # represents the G band of bulk hBN; (c) bare Ag particles on $SiO_2$ without BN protection (Ag/$SiO_2$) for 7 cycles; (d) ~1 nm thick ALD $Al_2O_3$ on Ag nanoparticles ($Al_2O_3$/Ag), and (e) 10 nm thick graphene on Ag nanoparticles (G/Ag) for 1 cycle, where * indicates the G band of graphite. (f) Summarized change of the Raman enhancement of these SERS substrates based on the intensity of the 612 $cm^{-1}$ peak. The black lines in (a) and (b) are Raman spectra of 2L/Ag and Bulk/Ag after heating at 360 °C in air for 5 min. The dashed lines for G/Ag in (e) and (f) mean the graphene completely oxidized and disappeared after the first cycle.

Atomically thin BN nanosheets also make SERS substrates highly reusable. Heating in air has been found the most effective and efficient way to remove adsorbed organic molecules for reusable SERS.[9a, 9b] As shown in our previous studies,[6c, 8] BN nanosheets are a promising barrier to protect metals because of their excellent thermal stability, high impermeability, and absence of galvanic corrosion. This means that the BN covered Ag-based SERS substrates can have a long "shelf-time" without loss of sensitivity due to oxidation, and more importantly, be regenerated by simply heating in air to burn off the adsorbed molecules and then reused. The reusability of 2L/Ag, Bulk/Ag and Ag/$SiO_2$ substrates for up to 30 cycles are shown in Figure 4a-c. In each cycle, the substrate was heated at 360 °C in air, and then reused by immersion in R6G solution ($10^{-6}$ M). The Raman signal and reusability of Ag/$SiO_2$ cannot be obtained at concentrations lower than $10^{-6}$ M. After the heating/cleaning treatment, no Raman signal of R6G was present on both 2L/Ag and Bulk/Ag samples, except a Si band at ~1000 $cm^{-1}$ and/or G band of hBN at 1366 $cm^{-1}$ (black spectra in Figure 4a and b), implying effective removal of the analyte. After 30 cycles of reusability tests, the Raman intensity and frequency of R6G from 2L/Ag remained essentially unchanged (Figure 4a), indicating that the Ag nanoparticles were not affected even after the repeated heating in air. In comparison, the Raman enhancement from Bulk/Ag substrate slightly weakened with the increase of cycle numbers (Figure 4b). This difference can be attributed to the different morphologies of atomically thin and bulk hBN on Ag nanoparticles. Atomically thin BN nanosheets are very flexible so that they wrinkle to better "seal" the underlying Ag nanoparticles from air; whereas thick hBN was unable to deform much, and there existed a gap between thick BN and $SiO_2$ substrate where air can penetrate to oxidize the Ag nanoparticles. Figure 2a-d, especially the lower diagrams, can be referred to visualize such difference. The Raman sensitivity of the bare Ag nanoparticles without BN protection, in contrast, plummeted in the reusability tests (Figure 4c). To better compare their different reusability, the Raman sensitivity of the three substrates after the tests are summarized in Figure 4f. The 2L/Ag and Bulk/Ag substrates retained ~90% and ~60% of their SERS activity after 30 cycles, respectively. For $Ag/SiO_2$, its plasmon enhancement decreased to 70% after the *first* cycle and further reduced to 20% after 7 cycles. Atomically thin BN also has a long-term protection on Ag nanoparticles: 2L/Ag kept 90% of its Raman enhancement after exposure to ambient condition for 1 year; while the $Ag/SiO_2$ totally lost its enhancement (Supporting Information, Figure S7).

For comparison, we also tested the reusability of two representative coatings, namely $Al_2O_3$ and graphene, as both materials have been proposed as protective barriers for SERS substrates with Ag nanoparticles.[4a, 10c] A 1.07 nm thick film (slightly thicker than 2L BN) of $Al_2O_3$ was grown on $Ag/SiO_2$ ($Al_2O_3/Ag$) by atomic layer deposition (ALD) (see Supporting Information, Figure S8), which can produce highly dense and uniform thin films.[17] $Al_2O_3/Ag$ showed Raman enhancement almost 10 times weaker than 2L/BN mainly due to two reasons. First, $Al_2O_3$ has a low affinity for R6G so that much less amounts of molecules can be adsorbed on the substrate. Second, it was reported before that a 1.5 nm thick $Al_2O_3$ film can reduce the Raman enhancement by more than 75%.[11] Moreover, the $Al_2O_3$ film showed almost no protection which could be due to inevitable defects in the film (Figure 4d), and its enhancement dropped to 20% after 7 cycles of reusability test, following a similar trend of $Ag/SiO_2$ (green line in Figure 4f).

High-quality graphene of 10 nm was produced on Ag nanoparticles by mechanical exfoliation, following the same synthesis route of BN/Ag. The SERS substrate of graphene covered Ag nanoparticles (G/Ag) showed comparable Raman enhancement to that of Bulk/Ag because graphene can also attract R6G via π-π interaction, but with lower background due to charge transfer (Figure 4e).[2a, 18] The G band of graphene (labelled by *) is pronounced compared to the peaks from R6G. The heating at 360 °C in air for 5 min (1 cycle) had disastrous effect on the Raman enhancement of G/Ag with a drop to 30% of the initial signal strength (blue dashed line in Figure 4f). It was because the ~10 nm thick graphene burnt out after the heating treatment (see Supporting Information, Figure S9), which dramatically decreased the amount of R6G molecules adsorbed on bare $Ag/SiO_2$ without graphene.

X-ray photoelectron spectroscopy (XPS) was used to analyze the oxidation levels of Ag nanoparticles on these substrates after reusability tests (Figure 5). Ag nanoparticles protected by BN nanosheets showed a dominating silver peak at 368.2 eV, indicating almost no increase of silver oxide after 30 cycles of reusability tests; while Ag particles without protection of BN or covered by the $Al_2O_3$ thin film showed a strong silver oxide peak at 367.6 eV,[19] suggesting ~50% contents of oxide after reusability tests. These results further confirm that BN nanosheets are an excellent material for reusable SERS devices.

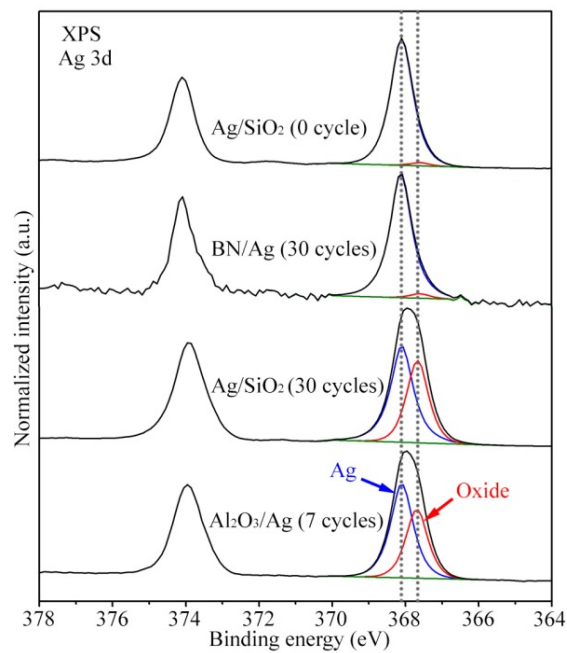

**Figure 5.** XPS spectra of Ag particles with and without BN protection before and after reusability tests.

In summary, atomically thin BN nanosheets can be used to fabricate highly sensitive, reproducible, and reusable SERS substrates. They are efficient in adsorption of aromatic molecules to improve Raman sensitivity especially at extremely low concentrations. The Raman enhancement from such substrates was homogeneous and reproducible. Furthermore, BN nanosheets could protect the Ag nanoparticles from oxidation even at high temperatures in air in the long term so that excellent reusability was achieved. The proposed SERS substrate and fabrication method can be easily scaled up using large-sized BN nanosheets grown by chemical vapor deposition (CVD).[20] On the other hand, BN nanosheets can be also used to cover SERS substrates produced from other methods, e.g. self-assembly, or other metal nanoparticles to improve surface adsorption, reproducibility, and reusability.

## Acknowledgements

L.H.Li thanks the financial support from the Australian Research Council under Discovery Early Career Researcher Award (DE160100796), and Deakin University under Alfred Deakin Research Postdoctoral Fellowship 2014, and Central Research Grants Scheme 2015. Y.Chen acknowledges the funding support from the Australian Research Council under the Discovery project. This work was performed in part at the Melbourne Centre for Nanofabrication (MCN).

**Keywords:** boron nitride nanosheets • surface enhanced Raman spectroscopy (SERS) • surface adsorption • reusability

[1] a) S. Schlucker, *Angew. Chem. Int. Ed.* **2014**, *53*, 4756-4795; b) P. H. C. Camargo, M. Rycenga, L. Au, Y. N. Xia, *Angew. Chem. Int. Ed.* **2009**, *48*, 2180-2184.

[2] a) X. Ling, L. M. Xie, Y. Fang, H. Xu, H. L. Zhang, J. Kong, M. S. Dresselhaus, J. Zhang, Z. F. Liu, *Nano Lett.* **2010**, *10*, 553-561; b) X. Ling, J. Zhang, *Small* **2010**, *6*, 2020-2025; c) H. H. Cheng, Y. Zhao, Y. Q. Fan, X. J. Xie, L. T. Qu, G. Q. Shi, *ACS Nano* **2012**, *6*, 2237-2244; d) Z. Y. Zhang, Q. H. Liu, D. L. Gao, D. Luo, Y. Niu, J. Yang, Y. Li, *Small* **2015**, *11*, 3000-3005.

[3] a) W. Xu, J. Xiao, Y. Chen, Y. Chen, X. Ling, J. Zhang, *Adv. Mater.* **2013**, *25*, 928-933; b) W. Xu, X. Ling, J. Xiao, M. S. Dresselhaus, J. Kong, H. Xu, Z. Liu, J. Zhang, *Proc. Natl. Acad. Sci. U.S.A.* **2012**, *109*, 9281-9286; c) Z. Zhang, F. G. Xu, W. S. Yang, M. Y. Guo, X. D. Wang, B. L. Zhanga, J. L. Tang, *Chem. Commun.* **2011**, *47*, 6440-6442; d) P. Luo, C. Li, G. Shi, *Phys. Chem. Chem. Phys.* **2012**, *14*, 7360-7366; e) J. Wang, X. Gao, H. Sun, B. Su, C. Gao, *Mater. Lett.* **2016**, *162*, 142-145.

[4] a) X. Li, J. Li, X. Zhou, Y. Ma, Z. Zheng, X. Duan, Y. Qu, *Carbon* **2014**, *66*, 713-719; b) M. Losurdo, I. Bergmair, B. Dastmalchi, T. H. Kim, M. M. Giangregroio, W. Y. Jiao, G. V. Bianco, A. S. Brown, K. Hingerl, G. Bruno, *Adv. Funct. Mater.* **2014**, *24*, 1864-1878.

[5] a) D. Prasai, J. C. Tuberquia, R. R. Harl, G. K. Jennings, K. I. Bolotin, *ACS Nano* **2012**, *6*, 1102-1108; b) S. Chen, L. Brown, M. Levendorf, W. Cai, S.-Y. Ju, J. Edgeworth, X. Li, C. W. Magnuson, A. Velamakanni, R. D. Piner, *ACS Nano* **2011**, *5*, 1321-1327.

[6] a) F. Zhou, Z. Li, G. J. Shenoy, L. Li, H. Liu, *ACS Nano* **2013**, *7*, 6939-6947; b) M. Schriver, W. Regan, W. J. Gannett, A. M. Zaniewski, M. F. Crommie, A. Zettl, *ACS Nano* **2013**, *7*, 5763-5768; c) L. H. Li, T. Xing, Y. Chen, R. Jones, *Adv. Mater. Interfaces* **2014**, *1*, 1300132.

[7] L. Liu, S. M. Ryu, M. R. Tomasik, E. Stolyarova, N. Jung, M. S. Hybertsen, M. L. Steigerwald, L. E. Brus, G. W. Flynn, *Nano Lett.* **2008**, *8*, 1965-1970.

[8] L. H. Li, J. Cervenka, K. Watanabe, T. Taniguchi, Y. Chen, *ACS Nano* **2014**, *8*, 1457-1462.

[9] a) P. Aldeanueva-Potel, E. Faoucher, R. A. Alvarez-Puebla, L. M. Liz-Marzan, M. Brust, *Anal. Chem.* **2009**, *81*, 9233-9238; b) Y. Lin, C. E. Bunker, K. S. Fernando, J. W. Connell, *ACS Appl. Mater. Interfaces* **2012**, *4*, 1110-1117; c) S. Faraji, K. L. Stano, O. Yildiz, A. Li, Y. Zhu, P. D. Bradford, *Nanoscale* **2015**, *7*, 17038-17047.

[10] a) X. Y. Zhang, J. Zhao, A. V. Whitney, J. W. Elam, R. P. Van Duyne, *J. Am. Chem. Soc.* **2006**, *128*, 10304-10309; b) J. F. Li, Y. F. Huang, Y. Ding, Z. L. Yang, S. B. Li, X. S. Zhou, F. R. Fan, W. Zhang, Z. Y. Zhou, D. Y. Wu, B. Ren, Z. L. Wang, Z. Q. Tian, *Nature* **2010**, *464*, 392-395; c) S. M. Mahurin, J. John, M. J. Sepaniak, S. Dai, *Appl. Spectrosc.* **2011**, *65*, 417-422; d) L. W. Ma, Y. Huang, M. J. Hou, Z. Xie, Z. J. Zhang, *Sci. Rep.* **2015**, *5*, 12890.

[11] J. A. Dieringer, A. D. McFarland, N. C. Shah, D. A. Stuart, A. V. Whitney, C. R. Yonzon, M. A. Young, X. Y. Zhang, R. P. Van Duyne, *Faraday Discuss.* **2006**, *132*, 9-26.

[12] a) X. Ling, W. Fang, Y.-H. Lee, P. T. Araujo, X. Zhang, J. F. Rodriguez-Nieva, Y. Lin, J. Zhang, J. Kong, M. S. Dresselhaus, *Nano Lett.* **2014**, *14*, 3033-3040; b) P. Dai, Y. Xue, X. Wang, Q. Weng, C. Zhang, X. Jiang, D. Tang, X. Wang, N. Kawamoto, Y. Ide, *Nanoscale* **2015**, *7*, 18992-18997.

[13] a) T. Taniguchi, K. Watanabe, *J. Cryst. Growth* **2007**, *303*, 525-529; b) L. Li, L. H. Li, Y. Chen, X. J. Dai, P. R. Lamb, B. M. Cheng, M. Y. Lin, X. Liu, *Angew. Chem. Int. Ed.* **2013**, *52*, 4212-4216.

[14] Q. Cai, L. H. Li, Y. Yu, Y. Liu, S. Huang, Y. Chen, K. Watanabe, T. Taniguchi, *Phys. Chem. Chem. Phys.* **2015**, *17*, 7761-7766.

[15] L. H. Li, Y. Chen, *Langmuir* **2010**, *26*, 5135-5140.

[16] L. H. Li, E. J. Santos, T. Xing, E. Cappelluti, R. Roldán, Y. Chen, K. Watanabe, T. Taniguchi, *Nano Lett.* **2014**, *15*, 218-223.

[17] M. Knez, N. Pinna, Wiley-VCH; John Wiley, Weinheim Chichester, **2011**.

[18] E. S. Thrall, A. C. Crowther, Z. H. Yu, L. E. Brus, *Nano Lett.* **2012**, *12*, 1571-1577.

[19] a) E. Albiter, M. Valenzuela, S. Alfaro, G. Valverde-Aguilar, F. Martínez-Pallares, *J. Saudi Chem. Soc.* **2015**, *19*, 563-573; b) E. Mysak, J. Smith, P. Ashby, J. Newberg, K. Wilson, H. Bluhm, *Phys. Chem. Chem. Phys.* **2011**, *13*, 7554-7564.

[20] a) Y. M. Shi, C. Hamsen, X. T. Jia, K. K. Kim, A. Reina, M. Hofmann, A. L. Hsu, K. Zhang, H. N. Li, Z. Y. Juang, M. S. Dresselhaus, L. J. Li, J. Kong, *Nano Lett.* **2010**, *10*, 4134-4139; b) L. Song, L. J. Ci, H. Lu, P. B. Sorokin, C. H. Jin, J. Ni, A. G. Kvashnin, D. G. Kvashnin, J. Lou, B. I. Yakobson, P. M. Ajayan, *Nano Lett.* **2010**, *10*, 3209-3215; c) Y. Gao, W. C. Ren, T. Ma, Z. B. Liu, Y. Zhang, W. B. Liu, L. P. Ma, X. L. Ma, H. M. Cheng, *ACS Nano* **2013**, *7*, 5199-5206.

# Supporting Information

## 1. Experimental details

**Adsorption test.** BN nanosheets were mechanically exfoliated from hBN single crystals on 90 nm thick $SiO_2$/Si substrate. The samples were heated at 350 °C in air for 1 h to remove tape residuals. Atomically thin BN was located under an Olympus optical microscopy (BX51) equipped with a DP71 camera, and the thickness was measured using a Cypher AFM (Asylum Research) in both tapping and contact modes. The BN samples were immersed in R6G aqueous (Milli-Q water) solution of $10^{-6}$ M for 1 min, then washed by water, and dried in argon (Ar) flow at room temperature. The percentages of coverage was estimated using the histogram function in Photoshop.

**Preparation of SERS substrates.** A ~10 nm thick layer of Ag film was deposited on $SiO_2$ (140 nm)/Si wafer by sputtering method under Ar (pressure of ~$5 \times 10^{-2}$ mbar) (SCD050, Bel-Tec). The current was 40 mA, and the sputtering time ranged from 20 to 40s. Then, BN nanosheets were mechanically exfoliated on the Ag film, following the aforementioned procedure. Subsequently, heat treatment at 450 °C in Ar for 1 h transferred the Ag film to particles of suitable sizes and distributions. The BN and graphene nanosheets were mechanically exfoliated following the same procedure from hBN single crystals and highly ordered pyrolytic graphite (HOPG), respectively. The SEM image of BN/Ag was taken using a Supra 55VP (Carl Zeiss) under 1.0 kV. The deposition of the $Al_2O_3$ coatings was carried out in a Fiji F200 (Cambridge Nanotech) ALD reactor using trimethyl aluminum (TMA) and water ($H_2O$) as precursors. Ar was used as a carrier and purging gas at a flow rate 20 sccm. The deposition temperature was 250 °C, and pressure was 200 mtorr. Each deposition cycle consisted of 0.02 s pulse of TMA, 10 s of Ar purge, 0.06 s pulse of $H_2O$, and finally 10 s Ar purge. For 1 nm coating, 11 cycles were conducted. The exact thickness of the deposited layer (1.07 nm) was determined by a spectroscopic ellipsometer using a blank Si wafer from the same ALD process.

**Raman analysis.** The BN veiled Ag particles on $SiO_2$/Si substrate was soaked in $10^{-8}$ M R6G (≥95%, Fluka) water solution for 1 h, washed with Milli-Q water to remove drops of R6G solution on the substrate, and then dried under Ar flow at room temperature. The reusability was tested by heating the substrate with R6G at 360 °C in air for 5 min and re-immersed in R6G solution of $10^-$ $^6$ M for 1 h for up to 30 times. Raman spectra were collected by a Renishaw Raman microscope equipped with a 514.5 nm laser. A 100× objective lens with a numerical aperture of 0.90 was used. The laser power was ~2.5 mW. In mapping, the step size was 0.4 μm. All Raman spectrum were calibrated with the Raman band of Si at 520.5 cm$^{-1}$.

**XPS analysis.** The XPS data were collected in high vacuum ($3 \times 10^{-9}$ torr) using a Kratos AXIS Ultra (DLD) spectrometer (Kratos Analytical Ltd, U.K.) equipped with a monochromated Al Kα X-ray source. All XPS spectra were recorded at 0.1 eV/step and a pass energy of 20 eV. For samples Ag/SiO$_2$ before and after 30 cycles of reusability tests and Al$_2$O$_3$/Ag after 7 cycles, the signals were collected from areas of $500 \times 300$ μm. For BN/Ag, the signals were from a square area of $27 \times 27$ μm due to the relatively small sizes of BN nanosheets. The location was determined by XPS mapping in N 1s region (Figure S1).

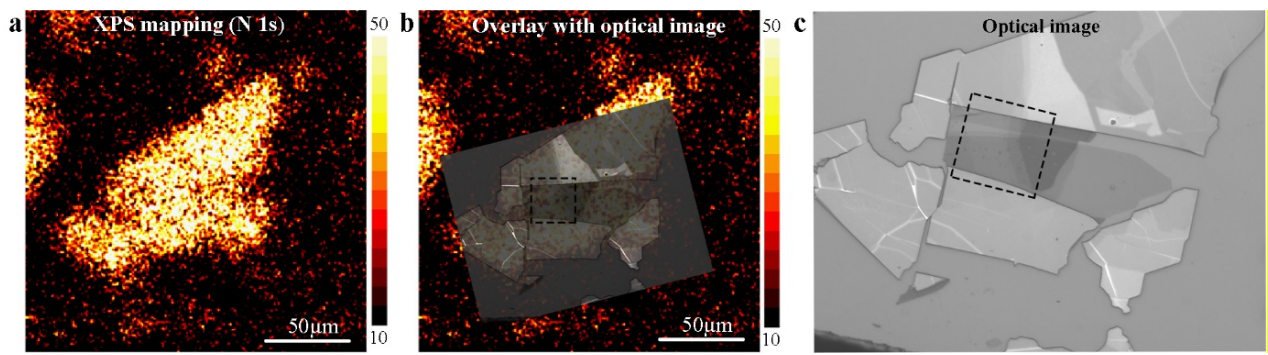

***Figure S1.*** *(a) XPS N1s mapping of BN nanosheets on Ag nanoparticles; (b) N1s mapping overlay with optical image; (c) optical image of BN on silver film. The XPS spectrum was collected form the square area ($27 \times 27$ μm) in (b) and (c).*

## 2. Optical microscopy images of BN nanosheets on Ag film and nanoparticles

Figure S2 shows the optical images of 1L, 2L, 17L BN nanosheets on Ag film before annealing (a) and on Ag nanoparticles after annealing (b). It can be seen that the optical contrast for atomically thin BN nanosheets became higher after annealing, and it was easier to locate them during SERS measurements.

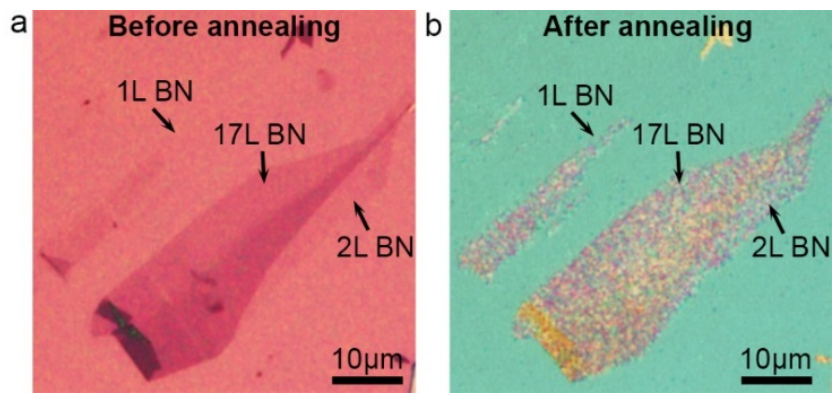

***Figure S2.*** *Optical microscopy images of exfoliated BN nanosheets on Ag/SiO₂/Si substrate before (a) and after (b) annealing in Ar gas at 450 ℃ for 1 h.*

## 3. Thickness of BN nanosheets

The thickness of the BN nanosheets was determined by AFM under contact mode before annealing when the Ag film was relatively flat (Figure S3). The measured thickness of the 2L BN is ~0.79 nm, and ~5.91 for the 17L BN.

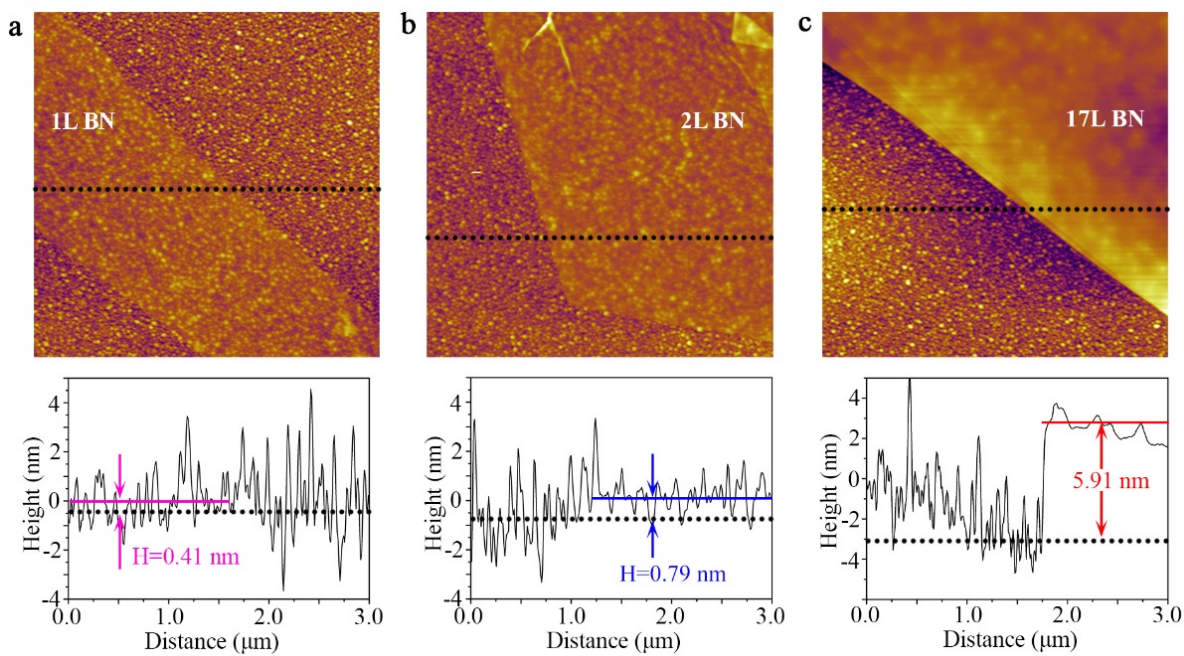

***Figure S3.*** *AFM images of 1L (a), 2L (b) and 17L BN (c) on Ag film and the corresponding height traces.*

## 4. Size distributions of Ag nanoparticles before and after reusability test

The size distributions of Ag particles on Ag/SiO$_2$, 2L/BN, Bulk BN, and Al$_2$O$_3$/Ag substrates were estimated from SEM images. Basically, the size of Ag particles retained unchanged after reusability tests (Figure S4), and hence no Ostwald ripening happen during heating for reusability tests.

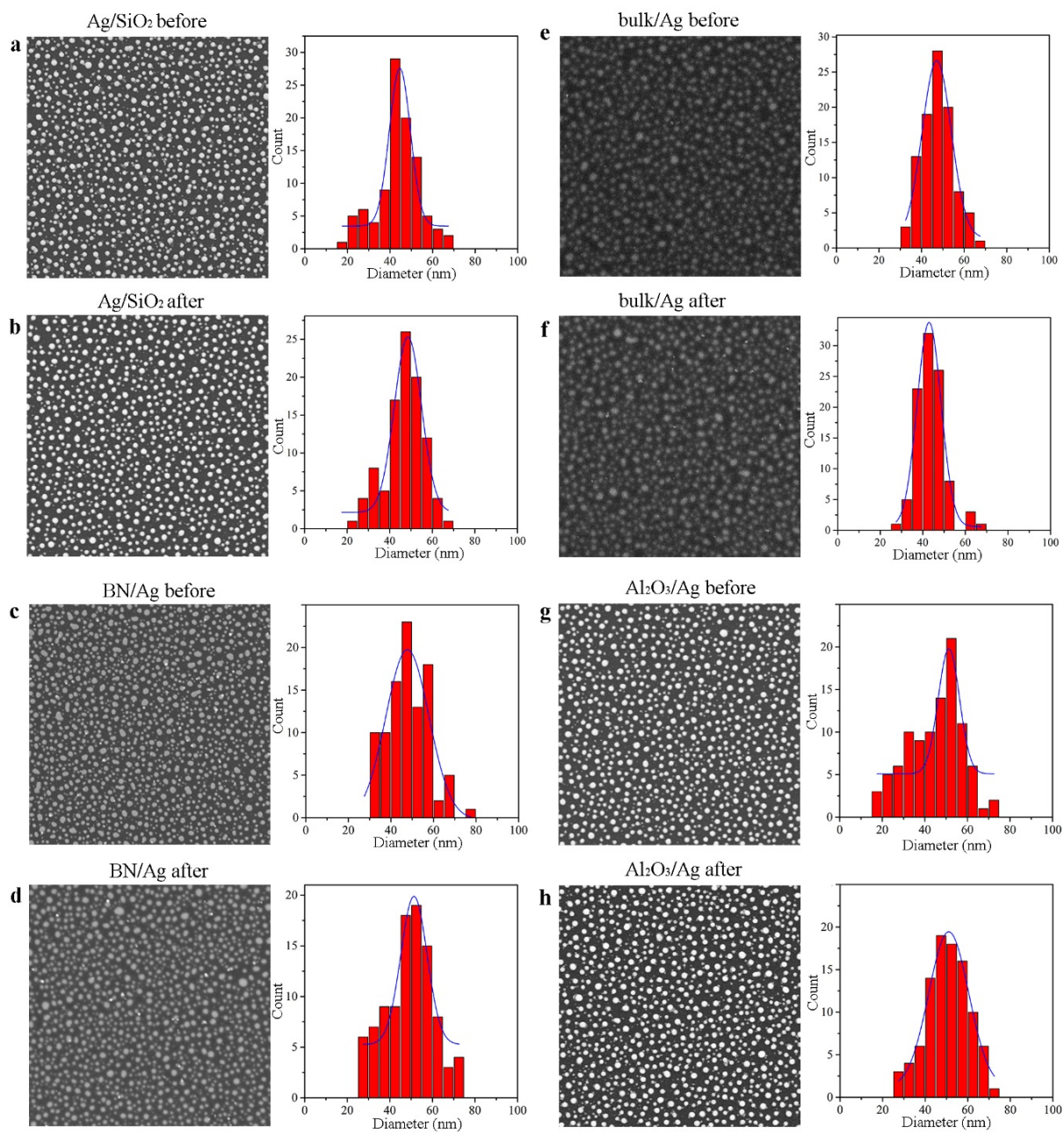

***Figure S4.*** *SEM images of Ag particles and the corresponding size distributions before and after reusability tests: (a) and (b) Ag/SiO₂; (c) and (d) BN/Ag; (e) and (f) Bulk/Ag; (g) and (h) Al₂O₃/Ag.*

## 5. Ag/SiO₂ substrates show no Raman signal of $10^{-7}$ M R6G

As shown in Figure S5, the typical Raman peak of R6G at 610 cm$^{-1}$ is undetectable on Ag/SiO₂ substrate when the concentration is $10^{-7}$ M.

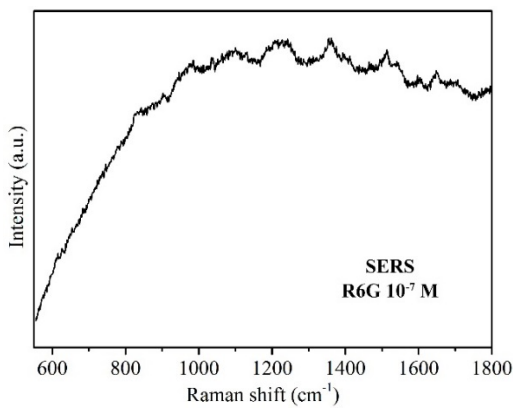

***Figure S5.*** *SERS spectra from Ag/SiO₂ substrate after immersion in R6G solution ($10^{-7}$ M) for 1 h.*

## 6. Enhancement factors

The much stronger Raman signals from 2L/Ag substrates should be mainly due to the superior surface adsorption of atomically thin BN towards R6G molecules. This was further confirmed by estimating the enhancement factor (EF) of the different substrates using droplet method to deposit R6G on substrates (1 droplet (20 µL) of R6G solution dripped on the substrates) because immersion method could not quantitively reveal their Raman enhancements. The EFs for Ag/SiO₂, 2L/Ag, and Bulk(0.5 µm thick BN)/Ag were close: $1.1×10^5$, $5.1×10^5$, and $0.9×10^5$, respectively. The slightly higher EF from atomically thin BN covered Ag substrate was due to the better surface adsorption of R6G on atomically thin BN due to π-π interactions, even though droplet instead of immersion method was used. That is, with much suppressed effect of surface adsorption, the enhancement by 2L/Ag was only 5 times of that by Ag/SiO₂. Therefore, BN nanosheets showed

SERS signals 2 orders higher than Ag/SiO$_2$ substrate can be mainly attributed to the effect of surface adsorption.

## 7. R6G's Raman background from BN/Ag as a function of laser energy

We tried 514.5, 632.8, and 785 nm lasers to test the effect of exciting laser energy on the background of R6G on BN/Ag substrate. In these tests, the substrate was immersed in R6G water solution (10$^{-6}$ M) for 1 h.

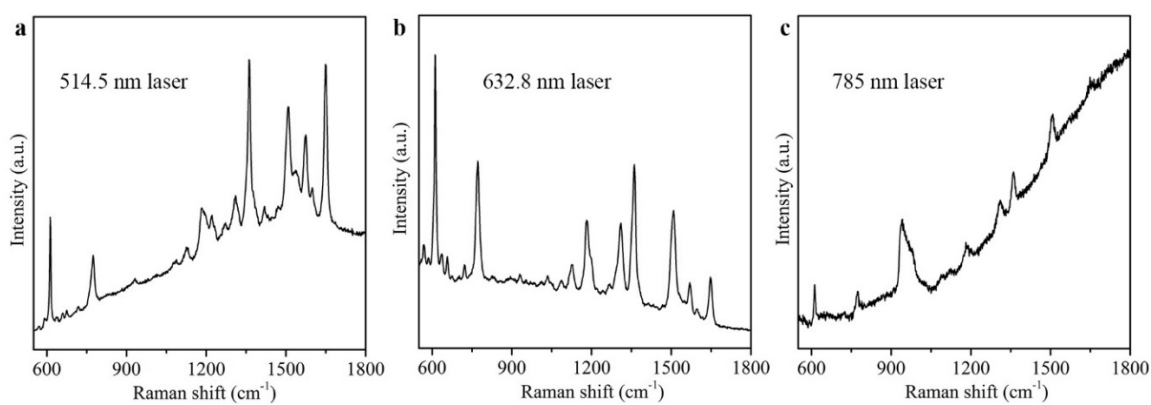

**Figure S6.** *Raman signals of R6G excited by lasers of different energies (a) 514.5 nm; (b) 632.8 nm; and (c) 785 nm.*

## 8. Long-term protection of Ag nanoparticles by 2L BN nanosheets

The protection of 2L BN on Ag nanoparticles was tested under ambient condition over a period of one year. It was found that 90% of the Raman enhancement from 2L/Ag substration was remained, whereas bare Ag/SiO$_2$ without BN protection totally lost Raman activity (Figure S7).

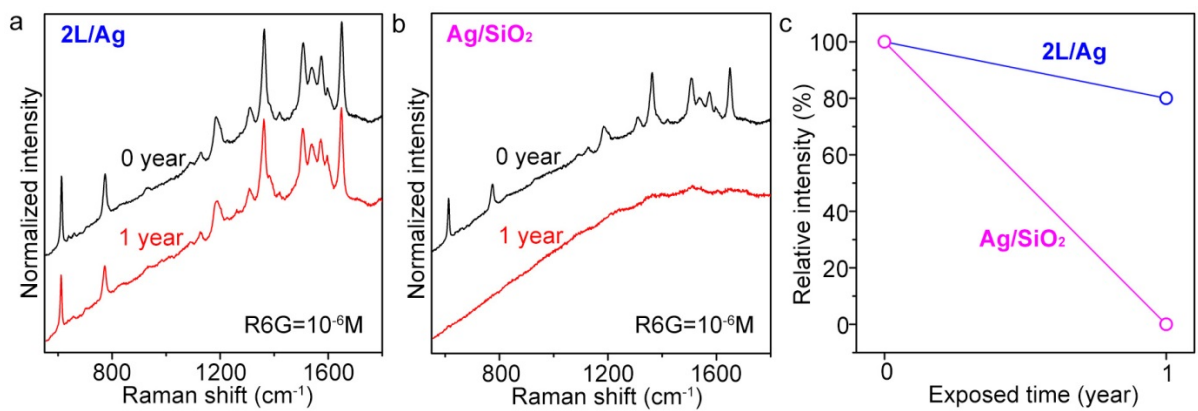

***Figure S7.*** *Raman enhancements of (a) 2L BN covered Ag nanoparticles and (b) bare Ag nanoparticles on SiO₂/Si without BN protection after exposed to ambient condition for 1 year. This was tested by immersio the substrates in 10⁻⁶ M R6G solution for 1 h (c) Summary of the change of Raman enhancement of the two substrates.*

## 9. AFM images of the ALD Al₂O₃ film

The 1.07 nm Al₂O₃ film was deposited on Ag particles and SiO₂/Si substrate by ALD method, and Figure S8 shows AFM images of the film at different magnifications.

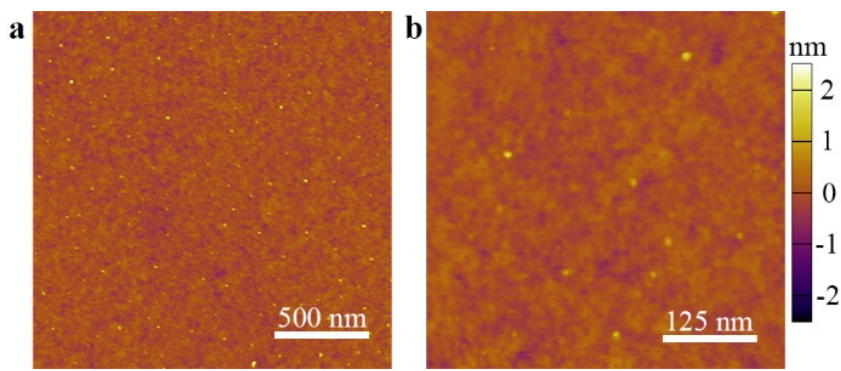

***Figure S8.*** *AFM images at different magnifications of the ~1 nm Al₂O₃ on SiO₂/Si substrate.*

## 10. Graphene burnt-out after 1 cycle of reusability test

Graphene was deposited on Ag film following the same procedure, and then the substrate was heated at 450 ºC in Ar for 1 h to allow the formation of Ag particles. The AFM image of the just exfoliated graphene on Ag film is shown in Figure S9a. After Raman measurement, the substrate was heated under the same condition for reusability test, i.e. at 360 ºC in air for 5 min. According to the AFM image in Figure S9b, it can be seen that the 10 nm thick graphite burnt out during the heating treatment of reusability. Therefore, graphene cannot protect Ag nanoparticles for reusable SERS substrate by heating.

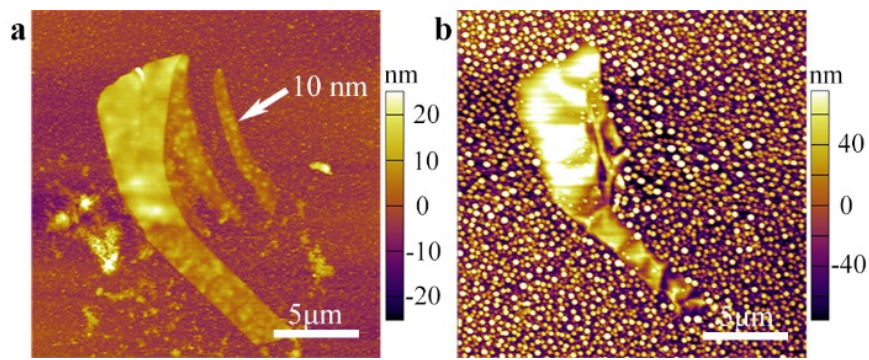

***Figure S9.*** *AFM images of (a) exfoliated graphene/graphite on Ag film, and (b) after 360 ˚C heating treatment in air for 5 min where the 10 nm-thick graphene disappeared.*